\documentclass[AER]{AEA}
\usepackage{makecell} 
\usepackage{booktabs}
\usepackage{multirow}
\usepackage{graphicx}
\usepackage{url}
\usepackage{float}  % in your preamble
\usepackage{placeins}
\usepackage{pdflscape}   % Landscape 
\usepackage{hyperref}

\usepackage{natbib}

\draftSpacing{1.5}

\begin{document}

\title{Killed in and after Action: The Long-lasting Effects of Combat Exposure on Mortality}
\shortTitle{Killed in and after Action}
\author{Helmut Farbmacher and Rebecca Groh\thanks{We thank Joshua Angrist for his helpful comments and Kevin Scotland and Napassakorn Atireklapwarodom for their excellent research assistance. Rebecca Groh gratefully acknowledges the MIT Economics Department for hosting her during a research stay, where part of this work was conducted. The replication code and instructions for obtaining the data are available at \url{https://github.com/farbmacher/KIA-and-KAA}. Corresponding author: Rebecca Groh: TUM School of Management, Technical University of Munich, Arcisstr. 21, 80333 Munich, Germany (rebecca.groh@tum.de). }} %Acknowledgements
\date{\today}
\pubMonth{Month}
\pubYear{Year}
\pubVolume{Vol}
\pubIssue{Issue}

\begin{abstract}

This study examines long-term mortality effects of combat exposure using the Vietnam War draft lottery as a quasi-experiment. We validate the lottery by analyzing combat fatalities, revealing that 1951-1952 cohorts had notably fewer lottery-induced deployments than 1950, limiting detectable long-term mortality impacts at the cohort level. Using deceased-only datasets, we invert standard identification by modeling draft eligibility as the outcome. We find significant excess mortality among Black men in the 1950 cohort (1.09\%, approximately 2,700 additional deaths), and null effects in later cohorts. Findings suggest that pooling cohorts with limited combat exposure may prevent detection of true treatment effects at cohort levels.

\end{abstract}

\maketitle

\iffalse
American Economic Review Pointers:

\begin{itemize}
\item Do not use an "Introduction" heading. Begin your introductory material before the first section heading.

\item Avoid style markup (except sparingly for emphasis).

\item Avoid using explicit vertical or horizontal space.

\item Captions are short and go below figures but above tables.

\item The tablenotes or figurenotes environments may be used below tables or figures, respectively, as demonstrated below.

\item If you are using an appendix, it goes last, after the bibliography. Use regular section headings to make the appendix headings.

\item If you are not using an appendix, you may delete the appendix command and sample appendix section heading.

\item Either the natbib package or the harvard package may be used with bibtex. To include one of these packages, uncomment the appropriate usepackage command above. Note: you can't use both packages at once or compile-time errors will result.

\item Manuscripts without exhibits may not exceed 7,000 words. Each exhibit reduces the maximum allowance by 200 words (e.g., a manuscript with five exhibits has a maximum of 6,000 words).

\item The maximum number of exhibits (figures or tables) is five. Individual tables and figures are limited to one page each.

\item An abstract of 100 or fewer words is required.
\end{itemize}
\fi

\noindent Understanding the long-term consequences of exposure to combat is of great relevance for evaluating the economic and social costs of armed conflict. As military operations continue worldwide, identifying whether combat exposure leads to persistent effects on health and mortality remains a key question with significant implications for understanding the full lifecycle costs of military service. The Vietnam War provides a valuable empirical setting to study these effects, as draft eligibility was determined through a randomized lottery that enables quasi-experimental analysis. This natural experiment allows us to examine the impact of a treatment where the treatment and control group differ only in being assigned different numbers based on their birthdays.

\noindent The Vietnam War draft lottery has been widely used to examine the short- and long-term effects of combat exposure. The quasi-experimental nature of the lottery allows for the investigation of the causal effect of military service on various outcomes. Yet the evidence remains mixed. Some studies find substantial adverse effects of military service \citep{hearst1986delayed, boehmer2004postservice, boscarino2006posttraumatic, autor2011battle, schlenger2015prospective, schmitz2016long, wang2025effects}, while others using similar identification strategies report no lasting differences in health or mortality \citep{angrist1996identification, dobkin2009health, angrist2010did, conley2012long, case2024education}. These conflicting results underscore the need for further evidence on how combat exposure translates into mortality risk across different demographic groups and cohorts.

\noindent We contribute to this literature by revisiting the long-term mortality effects of Vietnam War service using an inverted identification strategy motivated by the structure of our data. Since both of our datasets include only deceased individuals, we cannot model survival directly. Instead of comparing mortality rates between draft-eligible and draft-exempt men, we use draft eligibility itself as the outcome variable, similar to the reduced form regression in \cite{siminski2011long}. This allows us to test whether individuals born on draft-eligible days are overrepresented among the deceased. We implement this design using two complementary datasets. The first dataset, the Combat Area Casualties Dataset \citep{NARA_CombatCasualties}, contains Vietnam combat casualties, allowing us to assess whether draft eligibility translated into actual combat exposure. We test whether the share of casualties born on draft-eligible days exceeds the expected share under a uniform distribution of births within each calendar month. Under the assumption that combat mortality is independent of birth date, rejecting the associated test of equality implies that draft-eligible men were disproportionately deployed and exposed to combat.  The second dataset, the Berkeley Unified Numident Mortality Database (BUNMD) \citep{breen2022berkeley}, captures post-service deaths between 1975 and 2007, enabling us to test for long-term mortality effects using the inverted estimation approach. 

\noindent Our analysis of long-term mortality reveals significant increases among Black men in the 1950 cohort, but null effects in the 1951 and 1952 cohorts. This pattern suggests that, for the 1950 cohort, the number of draft-eligible men exposed to combat was large enough for mortality effects to emerge at the cohort level, whereas later cohorts had too few deployed men to generate detectable cohort-level effects. This aligns with the relatively few combat deaths we observe in the 1951 and 1952 cohorts. The concentration of effects among Black men, rather than among all draft-eligible individuals, suggests important heterogeneity in how combat exposure translated into long-term health consequences across demographic groups.

\noindent The remainder of this paper is structured as follows. Section \ref{Background_Identification} provides information on the draft lottery (\ref{Background}) and outlines our empirical framework (\ref{Identificationstrategy}). Section \ref{Killed in Action} introduces the dataset on combat casualties  (\ref{Data_KIA}), and validates the lottery mechanism (\ref{Validation}). Section \ref{Killed after Action} presents the data on long-term mortality (\ref{Data_KAA}) and the estimated long-term effects of combat exposure on mortality (\ref{Results}). Section \ref{Discussion} concludes with a discussion of the implications of our findings.

\section{Background and Empirical Framework} \label{Background_Identification}

\subsection{Background} \label{Background}

\noindent We exploit the quasi-experimental variation generated by the Vietnam War draft lottery system implemented in the United States in the early 1970s. Before 1970, the U.S relied on a decentralized draft system in which local boards determined call-up orders using various deferment criteria and local quotas. On December 1, 1969, the Selective Service System held the first national draft lottery to determine the induction order for men born between January 1, 1944, and December 31, 1950. The lottery assigned a random sequence number (1-366) to each calendar day, with lower numbers indicating higher draft priority. Men born on days assigned numbers 1–195 in 1970 (1–125 in 1971; 1–95 thereafter) were draft-eligible, while those with higher numbers were effectively exempt.

\noindent The lottery was introduced relatively late in the Vietnam War. By 1970, nearly 80\% of all U.S. combat deaths in Vietnam had already occurred. Although lotteries were conducted annually till 1975 for cohorts up to 1956, the last induction occurred in December 1972, and draft authority expired in 1973. This timing matters, as many individuals drafted via the lottery were never deployed, creating important variation in the relevance of lottery outcomes across cohorts.

\noindent For men born 1944–1949, the lottery's impact is ambiguous. These men were already draft-eligible under the pre-lottery system, and many had been inducted before December 1969. Consequently, in most datasets, it is difficult to distinguish whether their military service resulted from the lottery mechanism or earlier conscription. 

\noindent The 1950 cohort was the first whose draft status was determined solely by the lottery. These men turned 20 in 1970, making them draft-eligible when the lottery system was implemented and when U.S. ground combat involvement in Vietnam remained substantial. The 1951 cohort turned 20 in 1971, when U.S. troop levels had already begun decreasing from their 1968-1969 peak. The 1952 cohort reached draft age in 1972, by which time the U.S. had shifted to a policy of "Vietnamization" and was withdrawing combat forces. Although men from these cohorts could be formally drafted, the probability of actual combat deployment was substantially lower than for the 1950 cohort.

\noindent This institutional context suggests that the lottery provides the cleanest quasi-experimental variation for the 1950 cohort, where draft eligibility both determined conscription and translated into a meaningful risk of combat exposure.

% Mit Zitaten aus Helmuts Buch belegen

\subsection{Empirical Framework} \label{Identificationstrategy}
 
\begin{figure}[htbp]
\centering
	\includegraphics[width=0.75\textwidth]{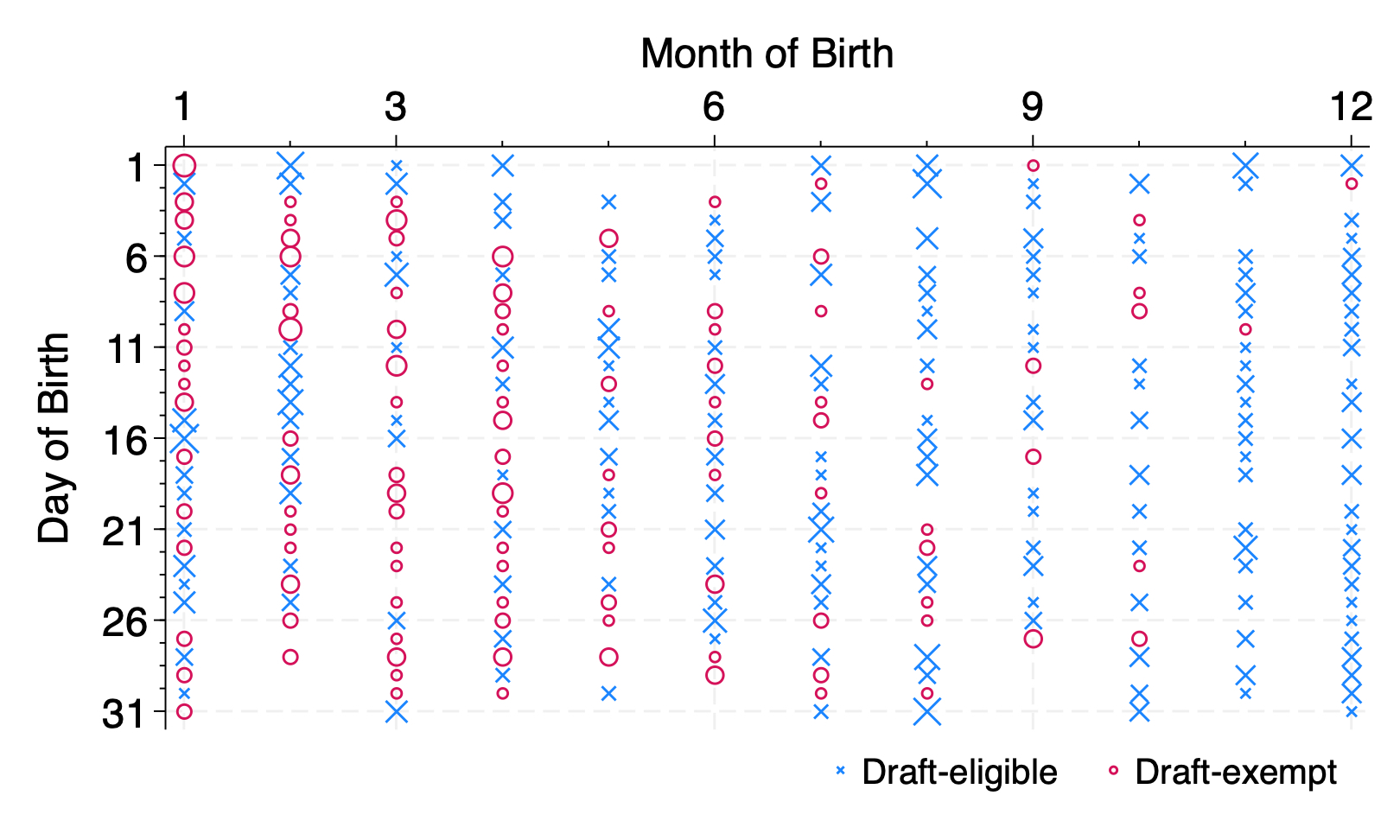} \\
	\includegraphics[width=0.75\textwidth]{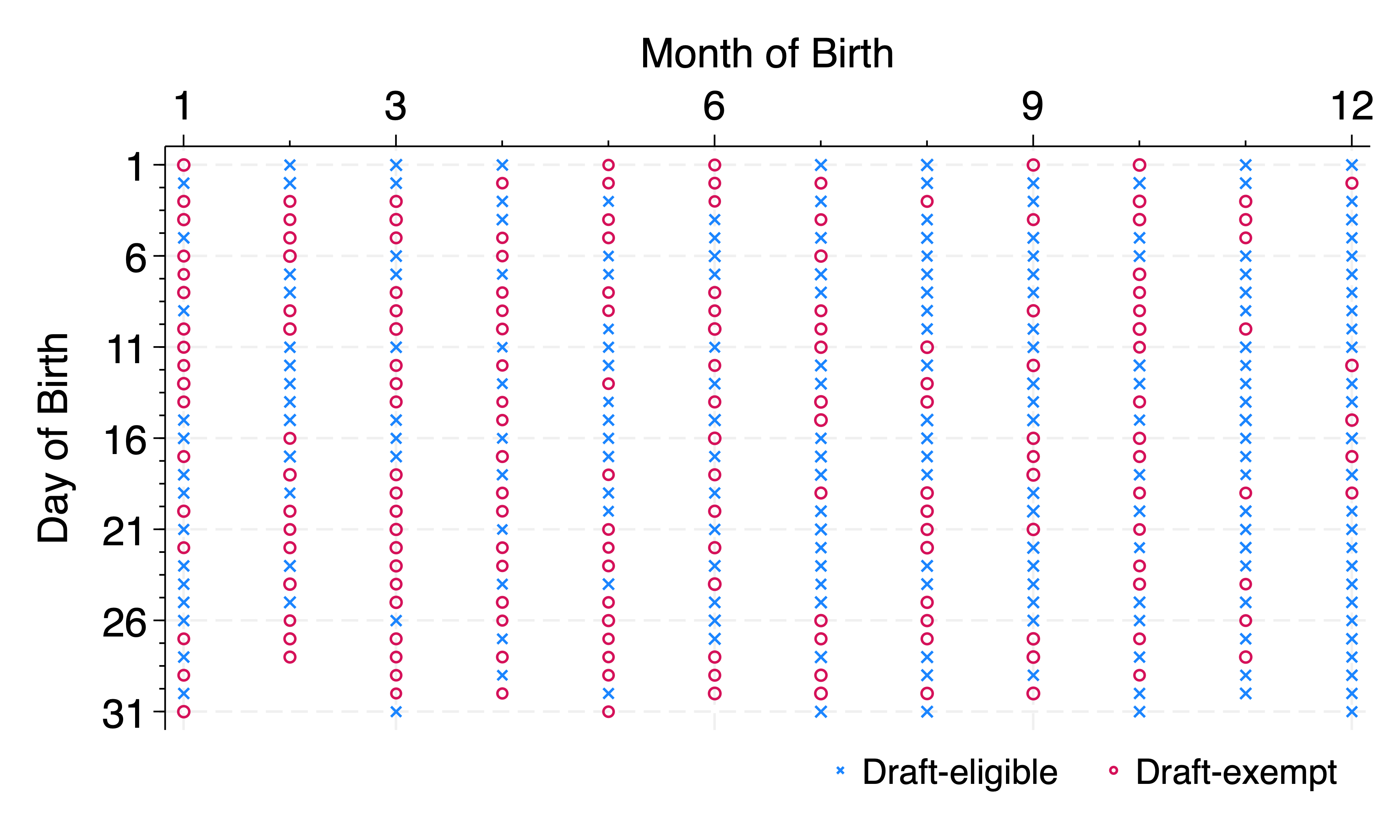}
\caption{Draft Lottery Assignment by Date of Birth in Both Datasets (1950 Cohort)}
\label{draft_days}
\begin{figurenotes} 
The upper panel illustrates the birthday distribution of draft-eligible and draft-exempt individuals from the 1950 cohort, using the Combat Area Casualties Dataset for Black and white Selective Service members who died before 1975. The lower panel displays the same distribution recorded in the BUNMD data (1975–2007) for all Black and white individuals, regardless of gender. Symbol size is proportional to the number of deaths for people born on that date in the respective dataset.
\end{figurenotes}
\end{figure}

\begin{figure}[h!]
\centering
\begin{minipage}[t]{0.49\textwidth}
    \centering
    \includegraphics[width=\textwidth]{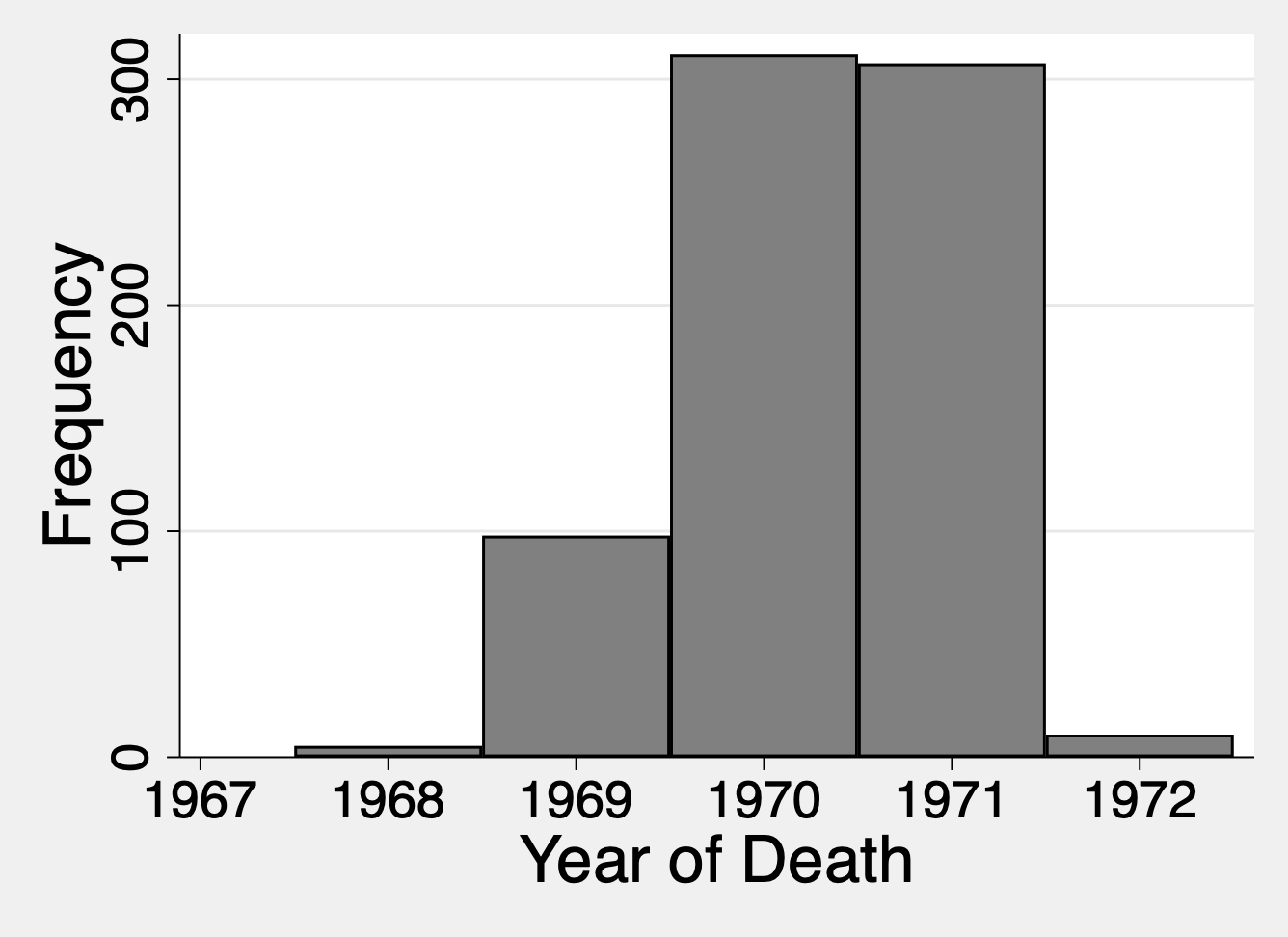}\\[2pt]
    Killed in Action
\end{minipage}\hfill
\begin{minipage}[t]{0.49\textwidth}
    \centering
    \includegraphics[width=\textwidth]{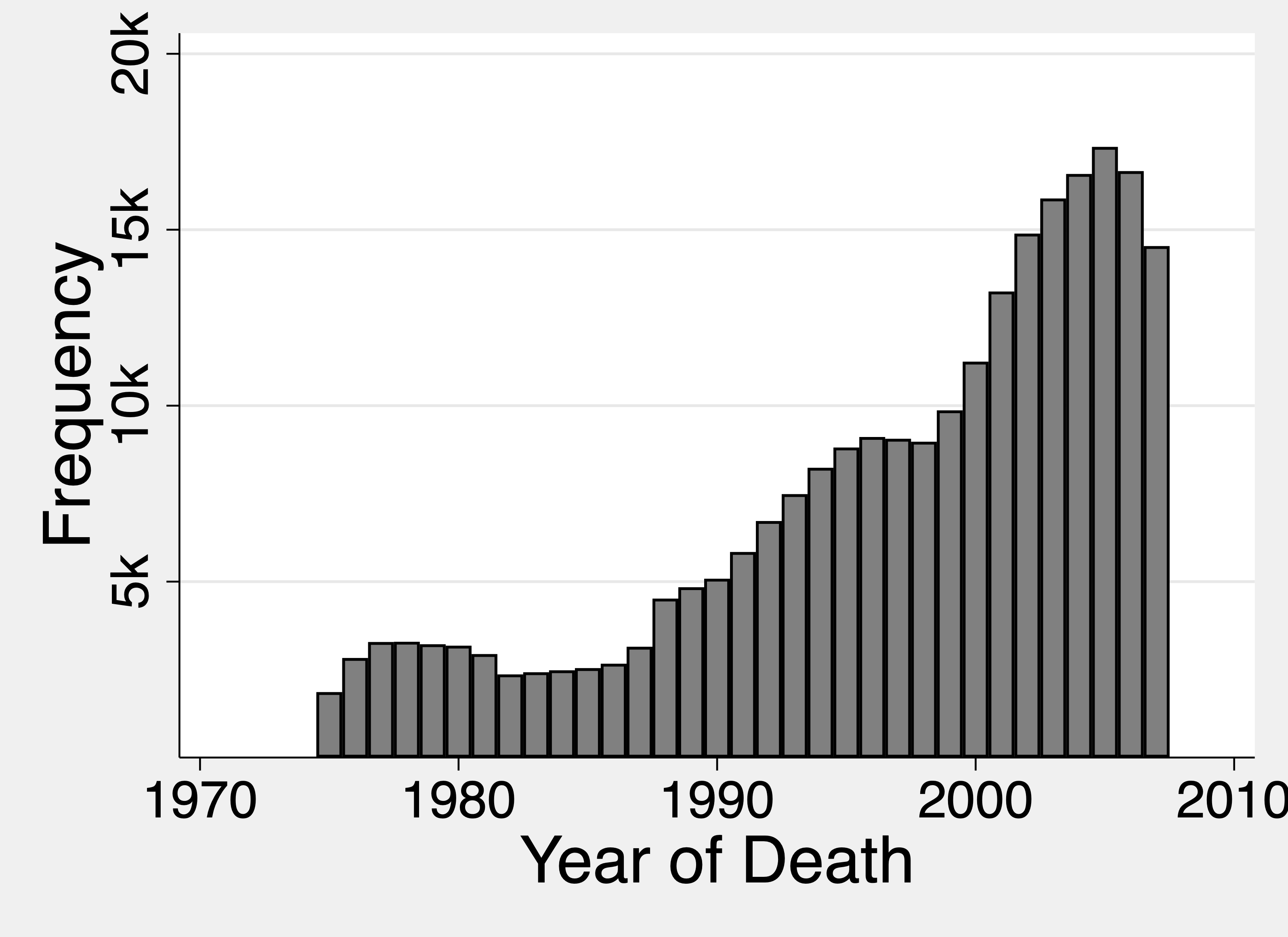}\\[2pt]
    Killed after Action 
\end{minipage}
\caption{Distribution of deaths for the 1950 cohort in the two datasets.}
\label{hist_deaths}
\begin{figurenotes}
The left panel displays the distribution of combat-related deaths for the 1950 birth cohort using the Combat Area Casualties Dataset, restricted to Black and white Selective Service members who died before 1975. The right panel shows the distribution of deaths for the same 1950 cohort as recorded in the BUNMD data (1975–2007), including all Black and white individuals, regardless of gender, who died during this period.
\end{figurenotes}
\end{figure}

\noindent We draw on two distinct datasets. The first, the Combat Area Casualties Dataset, which records U.S. soldier casualties during the Vietnam War, allows us to validate the lottery mechanism. The second, the BUNMD, which includes deaths of U.S. citizens between 1975 and 2007, is used to assess the long-term lifecycle costs of combat exposure. Figure \ref{draft_days} presents the distribution of draft-eligible and draft-exempt birthdays for the 1950 birth cohort in each dataset, showing that our data capture deaths across the full range of lottery assignments. Figure \ref{hist_deaths} provides corresponding histograms of deaths for the 1950 cohort.

\FloatBarrier

\noindent 

\FloatBarrier

\noindent To leverage the quasi-experimental setting of the U.S. draft lottery, our first step is to validate that the lottery mechanism functioned as intended. Specifically, we test whether men born on draft-eligible days were, in fact, more likely to be deployed to Vietnam and therefore exposed to combat, compared to men born on draft-exempt days. Because individual draft status is unobserved, combat deaths serve as a proxy for exposure. Assuming birthdays are uniformly distributed within calendar months and that the probability of being killed in action is independent of date of birth, any overrepresentation of combat deaths of men born on draft-eligible days indicates that the lottery effectively increased deployment and combat exposure. The expected fraction of combat deaths among draft-eligible men is calculated by multiplying the total number of male births in a given month (by cohort and race, from Vital Statistics 1944–1953) by the fraction of days classified as draft-eligible. We then compare this expected fraction to the observed fraction.
  
\noindent After validating the impact of the lottery on combat exposure, we estimate the long-term lifecycle costs measured by excess mortality. Because the BUNMD datasets consist solely of deceased individuals, we cannot use survival as the outcome variable. Instead, we analyze the probability of being born on a draft-eligible day among the deceased, using draft-eligibility as the outcome variable. This approach allows us to test whether deaths are overrepresented for certain groups among those who are draft-eligible, indicating excess mortality associated with combat exposure. We use the following regression specification:

\begin{equation}
\text{Draft-eligibility}_i = \beta_0 
  + \beta_1 \,\text{Black}_i
  + X_i' \gamma
  + \varepsilon_i
\end{equation}

\noindent  where $i$ indexes individual observations, $Draft$-$eligibility_i$ represents a binary indicator of draft-eligibility (1) or draft-exemption (0), and $Black_i$ is a binary indicator of the individual being Black. The vector $X_i$ captures month of birth fixed effects. 

\noindent We then extend the model to also include women, who were not subject to the draft, as a control group. Since draft eligibility was randomly assigned by date of birth, men and women should have identical probabilities of being born on draft-eligible days. This argument also holds true for other group differences. We estimate the following regression equation, now also including $Men_i$, which is a binary indicator of the deceased being male: 

\begin{equation}
\text{Draft-eligibility}_i = \beta_0 
  + \beta_1 \,\text{Black}_i
  + \beta_2 \,\text{Men}_i
  + \beta_3 \,\text{Black}_i \times \text{Men}_i
  + X_i' \gamma
  + \varepsilon_i
\end{equation}

\noindent These regressions allow us to test whether deaths are disproportionately concentrated among those born on draft-eligible days, indicating excess mortality. A positive and significant $\beta$ coefficient suggests that the respective group, Black individuals, men, or Black men, is overrepresented among the deceased due to draft-induced combat exposure.

\section{Killed in Action} \label{Killed in Action}

\subsection{Data} \label{Data_KIA}

\noindent We use the Combat Area Casualties Dataset, which contains records of U.S. military soldiers who died in the Southeast Asian combat area during the Vietnam War, either as a result of combat or following from injuries received in combat. The dataset covers a total of 58,193 deceased soldiers with deaths occurring between 1956 and 1998. It contains detailed information on date of birth, date of death, and race, among others \citep{NARA_CombatCasualties}. 

\noindent We concentrate our analyses on the draft-affected cohorts from 1944 to 1952 (13,708 observations dropped) and on individuals marked as Selective Service members, as they cover the lottery draftees, which leaves us with 16,056 observations. Further, we only include Black and white individuals (109 observations dropped) and individuals who died before 1975 (44 observations dropped), as later deaths could already be covered in the BUNMD dataset. We assign the lottery numbers for each respective birth cohort based on their date of birth. This enables us to differentiate between draft-eligible and draft-exempt men among the deceased. 

\noindent The presence of draft-exempt individuals among the deceased is expected for earlier cohorts (1944–1949), when conscription reflected a mix of local draft board procedures and lottery assignment, the later cohorts were drafted exclusively through the lottery system. Thus, the presence of draft-exempt individuals among the deceased likely reflects two mechanisms. First, some men may have already enlisted before the lottery was introduced. Second, draft-motivated enlistment is plausible, as the cutoff number for the 1950 cohort (195) was not announced until later in 1970 \citep[see][]{angrist1991draft}. Therefore, individuals who did not yet know their draft risk may have chosen to enlist preemptively in anticipation of being drafted.

\noindent To calculate the expected fraction of draft-eligible men in each cohort, we use the Vital Birth Statistics \citep{NCHS_VSUS}, which report monthly U.S. births by sex and race. In the 1950 cohort, there were 1,823,555 male births (51.31\% of all births); in the 1951 cohort, 1,923,020 male births (51.27\%); and in the 1952 cohort, 1,971,262 male births (51.24\%). 

\noindent The Vital Birth Statistics provide monthly birth counts in two categories, one for white individuals and another that combines all nonwhite individuals. More detailed racial information is available only on an annual basis. The data reveal that Black births comprise approximately 95\% of nonwhite births across the 1944-1952 cohorts, with percentages ranging from 93.87\% in 1944 to 95.68\% in 1947. Consequently, using nonwhite rather than Black monthly birth counts to calculate expected fractions of draft-eligible men provides a very close approximation.

\subsection{Results} \label{Validation}

\noindent We now present the results of the validation analysis, which tests whether draft eligibility translated into higher rates of exposure to combat. Specifically, we compare the share of combat deaths among draft-eligible individuals to the expected share. A statistically significant excess of draft-eligible deaths would indicate that the lottery mechanism effectively assigned higher combat risk to those selected.

\noindent Table \ref{fatal_casualties_all} presents results for all cohorts subject to the lottery draft system. For earlier cohorts not selected solely through the lottery (1944–1949), the observed and expected fractions of draft-eligible combat deaths do not differ significantly. This pattern is consistent with the historical context, since individuals from these cohorts were subject to conventional conscription before the lottery was introduced, so their exposure to combat cannot be clearly linked to lottery outcomes. In contrast, for the later cohorts drafted exclusively through the lottery (1950–1952), the 1950 and 1951 cohorts display statistically significant excesses of combat deaths for men born on draft-eligible days. For the 1950 cohort, 72.23\% of combat deaths occurred among men born on draft-eligible days, compared to an expected 53.77\%, yielding a difference of 18.46 percentage points ($t = 11.14$). The 1951 cohort also shows significant difference, with 60.92\% of combat deaths among draft-eligible men compared to an expected 34.25\% ($t = 5.07$). There is no significant difference in the 1952 cohort, which records very few fatalities, reflecting the declining intensity of U.S. involvement in Vietnam by the time these men were eligible for deployment.

\begin{table}[htbp]
\caption{Fatal Observed and Expected Casualties of Selective Service Soldiers, Cohorts 1944-1952}
\label{fatal_casualties_all}
\centering
\begin{tabular}{lrrrr}
\toprule
\multirow{2}{*}{\textbf{Cohort}} &
\multirow{2}{*}{\begin{tabular}[c]{@{}c@{}}\textbf{Total}\\\textbf{casualties}\end{tabular}} &
\multicolumn{2}{c}{\textbf{Born on draft-eligible days}} &
\multirow{2}{*}{\textbf{$t$-statistic}} \\
\cmidrule(lr){3-4}
& & \textbf{Observed} & \textbf{Expected} & \\
\midrule
1944 & 1111 & 616 (55.45\%) & 595.61 (53.61\%) & 1.23 \\
1945 & 1586 & 867 (54.67\%) & 846.92 (53.40\%) & 1.01 \\
1946 & 2762 & 1546 (55.97\%) & 1530.15 (55.40\%) & 0.61 \\
1947 & 4251 & 2279 (53.61\%) & 2251.75 (52.97\%) & 0.84 \\
1948 & 3461 & 1835 (53.02\%) & 1851.29 (53.49\%) & -0.55 \\
1949 & 1907 & 996 (52.23\%) & 1021.96 (53.59\%) & -1.19 \\
1950 & 731 & 528 (72.23\%) & 393.06 (53.77\%) & 11.14 \\
1951 & 87 & 53 (60.92\%) & 29.80 (34.25\%) & 5.07 \\
1952 & 7 & 3 (42.86\%) & 1.82 (25.94\%) & 0.84 \\
\bottomrule
\end{tabular}
\begin{tablenotes}
\small
The table reports the total number of Vietnam combat deaths by birth cohort and compares observed and expected counts for men born on draft-eligible days (195 draft-eligible days for 1944–1950 births, 125 in 1951, and 95 in 1952). Observed: reports the actual number of deaths among men born on draft-eligible days and their percentage of total cohort casualties. Expected: reports the predicted number and percentage based on the proportion of days in each cohort's birth year classified as draft-eligible (195 in 1944-1950, 125 in 1951, 95 in 1952), assuming uniform distribution of births across days in each month. The $t$-statistics test equality between observed and expected proportions.
\end{tablenotes}
\end{table}
 
 \FloatBarrier

\noindent We now focus on the 1950 cohort, for which the lottery mechanism appears to have operated most effectively. For Table \ref{fatal_casualties}, we further restrict the dataset by excluding individuals who died before 1970, as they could not have been drafted through the lottery. This reduces the dataset to a total of 628 combat deaths, of which 79 are Black and 549 are white individuals. 

\noindent Comparing the cohort’s total death records to the expected values, we find even larger differences (22.82 percentage points) than in Table \ref{fatal_casualties_all}, with the t-statistic increasing to $t = 13.50$. Analyzing the data by race shows the same overrepresentation of men born on draft-eligible days among both Black (28.42 percentage points; $t = 6.57$) and white men (22.01 percentage points; $t = 12.03$).  These results indicate that draft eligibility significantly increased the likelihood of combat exposure for both groups.

\begin{table}[htbp]
\caption{Fatal Observed and Expected Casualties of Selective Service Soldiers born 1950}
\label{fatal_casualties}
\centering
\begin{tabular}{lrrr}
\toprule
& \textbf{Overall} & \textbf{Black} & \textbf{White} \\ 
\midrule
\textbf{Total casualties} & 628 & 79 & 549 \\
\midrule
\multicolumn{4}{l}{\textbf{Born on draft-eligible days}} \\
\quad Observed & 481 (76.59\%) & 65 (82.28\%) & 416 (75.77\%) \\
\quad Expected & 337.68 (53.77\%) & 42.55 (53.86\%) & 295.14 (53.76\%) \\
\quad \textit{Difference} & \textit{+22.82 pp} & \textit{+28.42 pp} & \textit{+22.01 pp} \\
\midrule
$t$-statistic & 13.50 & 6.57 & 12.03 \\
\bottomrule
\end{tabular}
\begin{tablenotes}
The table reports the total number of Vietnam combat deaths (1970-1974) in the 1950 cohort and compares observed and expected counts for men born on draft-eligible days. Observed: reports the actual number of deaths among men born on draft-eligible days and their percentage of total cohort casualties. Expected: reports the predicted number and percentage based on the proportion of days classified as draft-eligible, assuming uniform distribution of births across days in each month. Vital Birth Statistics (expected values) provide only month-specific birth numbers for white and nonwhite individuals; since approximately 95\% of nonwhite individuals are Black, we directly compare expected fractions based on nonwhite births to observed Black casualties. The $t$-statistics test equality between observed and expected proportions.
\end{tablenotes}
\end{table}

\FloatBarrier

\noindent While long-term health and mortality effects are expected for war veterans in each birth cohort, the number of individuals actually exposed to combat declines sharply across cohorts. For the 1951 and 1952 cohorts, casualty numbers drop substantially, indicating that although the lottery continued to determine draft eligibility, far fewer draft-eligible men were ultimately deployed to combat. As a result, the impact of combat exposure may not be detectable at the cohort level in later cohorts, even if individual-level effects exist. Pooling cohorts with such heterogeneous deployment levels may dilute the treatment effect and reduce statistical power to detect long-term mortality impacts. Accordingly, the 1950 cohort represents the setting where draft eligibility through the lottery most clearly translated into combat exposure and thus provides the most credible basis for analyzing long-term mortality effects.

%\noindent Next, we conduct a cohort-specific regression to assess to what extent the immediate impact of combat exposure diﬀers between Black and white soldiers. The results show a highly significant, positive effect for Black men in the 1950 cohort, controlling for month-of-birth fixed effects ($\beta = 0.1127$, $SE = 0.0415$). This indicates that, within this cohort, draft-eligibility was associated with higher observed immediate mortality for Black men. In the next section we aim to figure out whether the increased combat exposure also accumulates to long-term mortality eﬀects.

\section{Killed after Action} \label{Killed after Action}

\subsection{Data} \label{Data_KAA}

\noindent We use the BUNMD dataset, a detailed version of administrative mortality records containing over 49 million entries for the years 1975 to 2007. The data includes information like date of birth, date of death, race, and sex. It has a high coverage of fatalities for the period from 1988 to 2005, capturing between 75\% and 90\% of all deaths among U.S. citizens \citep{breen2022berkeley}. This enables us to track deaths among the draft lottery–affected cohorts as they age from their twenties into their sixties.

\noindent For our analysis, we restrict the dataset to the draft-affected cohorts from 1944 to 1952, which leaves us with 2,779,988 observations, with primary attention to those born in 1950. We only include Black and white individuals in these cohorts (dropping 13.13 \% of the observations). Additionally, we exclude all observations lacking information on date of birth or gender to ensure accurate determination of the draft status and a clean sample construction (dropping 0.02 \% of the observations). The 1950 cohort includes more than 245,000 observations; therefore, people who died between 1975 and 2007, with 33.52\% women and 21.72\% Black individuals. The 1951 and 1952 data, respectively, contain more than 242,000 and more than 236,000 observations, with comparable shares of women and Black individuals. The dataset allows us to assess long-term excess mortality among Vietnam veterans, even though we are not able to observe who actually served in the military, using draft eligibility as a proxy. We assign the draft lottery number to each respective cohort based on the date of birth, also for women to serve as a control group. 

\noindent Since we are interested in excess mortality, we must assess whether our deceased-only dataset is prone to potential selection bias. If individuals exposed to combat had already died in excess before 1975, they would be underrepresented among the deceased individuals. To evaluate this concern, we compare the observed fraction of individuals born on draft-eligible days in the BUNMD to the expected fraction based on the underlying birth-date distribution in the population. 

\noindent For the 1950 cohort, the expected fraction of individuals born on draft-eligible days is 53.77 \% (identical when restricting to men). In the BUNMD data, the corresponding shares are 53.75\% overall and 53.74\% for men. We find the same pattern for the 1951 and 1952 cohorts. This absence of measurable selection is not inconsistent with evidence that combat exposure increased short-term mortality, see \citep{hearst1986delayed}. Instead, it implies that any such mortality differences may not be sufficiently large to be visible at the cohort level.

\subsection{Results} \label{Results}

\noindent Our cohort-specific analyses reveal heterogeneous long-term mortality effects. Table \ref{reg_results} reports the regression estimates for the 1950 cohort, where month-of-birth fixed effects are included to account for potential seasonality in birth patterns.

\noindent We first estimate the model using only male observations. The results indicate that Black men are disproportionally represented among the deceased born on draft-eligible days, although the coefficient is statistically significant only at the 10 \% level. 

\noindent When we extend the analyses to the combined sample of men and women, the individual effects of being Black or male are neither positive nor statistically significant. However, the interaction term between being Black and male is positive and statistically significant ($\beta = 0.0109$, $SE = 0.0049$), indicating that Black men born on draft-eligible days are 1.09 percentage points more likely to appear among the deceased compared to what would be expected under random assignment. This result suggests that Black men experience higher long-term mortality. With 248,246 Black men born in 1950, this excess mortality translates to roughly 2706 additional deaths attributable to draft-induced combat exposure over the 33-year follow-up period. 

\noindent In contrast, for the 1951 and 1952 cohorts, none of the regressors or interaction terms show statistically significant effects. The full estimation results for the 1944–1952 cohorts are presented in Table \ref{reg_results_multi} in the Appendix. Excluding the 1950 cohort, this table reports the results of 32 tests, of which only one, the effect of being Black in the 1947 cohort, is significant at the 5\% level, which is roughly the number one would expect by chance.

\begin{table}[htbp]
\caption{Effect of draft eligibility on long-term mortality for the 1950 cohort}
\label{reg_results}
\centering
\begin{tabular}{lcc}
\toprule
                        & \textbf{Men only} & \textbf{Men \& women} \\
\midrule
		$Black$ 			& 0.0051    & -0.0058  \\
							& (0.0029)  & (0.0040) \\
		$Men$ 				& 			& -0.0027  \\
							& 			& (0.0023) \\
		$Black \times men$	& 			& 0.0109   \\
							& 			& (0.0049) \\ \hline
		Sample size	 		& 162,974	& 245,151 \\
\bottomrule
\end{tabular}
\begin{tablenotes}
Estimates from OLS regression with controls for month of birth fixed effects. Robust standard errors are reported in parentheses.
\end{tablenotes}
\end{table}

\FloatBarrier

\noindent These findings support two key conclusions. First, there is indeed a measurable long-term mortality effect of combat exposure for Black men in the 1950 cohort. Second, while later cohorts experienced some level of combat exposure, the sample sizes of those actually exposed to combat are insufficient to generate detectable long-term mortality effects when examining the entire cohort.

\section{Discussion and Conclusion} \label{Discussion}

\noindent This study revisits the lifecycle costs of exposure to combat during the Vietnam War, using the U.S. draft lottery as a quasi-experimental setting. We contribute by validating whether draft eligibility effectively translated into combat exposure across cohorts and by applying an inverted identification strategy suited for deceased-only datasets.

\noindent We validate the lottery mechanism by comparing expected and observed exposure to combat, measured in terms of deaths as a proxy. The analyses show that the lottery mechanism was effective, particularly for the 1950 and 1951 cohorts, as we see significant differences. But the absolute numbers of combat exposure, which are consistent with the historical drawdown of U.S. troops and the implementation of “Vietnamization”, indicate that the sample size might be too small to translate into effects that are detectable at the cohort level, at least for the 1951 and 1952 cohorts.
% Mit Zitat von Helmuts Buch belegen 

\noindent Our results highlight the importance of aligning the empirical strategy with the institutional and historical context of the Vietnam War. For this reason, we conduct cohort-specific analyses, as pooling cohorts with differing exposure intensities may obscure or even eliminate the detection of true treatment effects.
% Mit Zitat von Helmuts Buch belegen 

\noindent Using the BUNMD dataset, we find significant excess mortality among Black men born on draft-eligible days.  Our findings contribute to a growing body of literature suggesting that the lifecycle costs of combat exposure extend well beyond the combat zone and are not evenly distributed across different demographics. The elevated long-term mortality among Black men underscores the compounding effects of racial inequality and military service.

\noindent The observed racial heterogeneity is consistent with prior research suggesting that Black soldiers faced greater or more hazardous exposure during the Vietnam War and experienced unequal access to postwar healthcare. Studies such as \cite{dismuke2015racial} show that Black veterans were disproportionately affected by mild traumatic brain injury when exposed to combat, compared to white veterans. Similarly, \cite{dohrenwend2008war}, \cite{penk1989ethnicity}, and \cite{allen1986posttraumatic} document higher rates of chronic PTSD among Black Vietnam veterans, attributing these disparities to greater or more severe exposure to war-zone stressors, racism within the military, and limited postwar opportunities. \cite{roberts2011race} further finds that Black veterans were less likely to receive treatment for PTSD compared to white veterans. These findings reinforce the view that the lifecycle costs of combat exposure were not evenly distributed across demographic groups.

\noindent While our study leverages two complementary datasets to analyze the draft lottery and the long-term mortality, it faces several limitations. First, our mortality data include only deceased individuals, which constrains the estimation framework and precludes direct modeling of survival. Second, we rely on draft eligibility as a proxy for combat exposure, since individual-level service records are unavailable in the death registers. Although our validation results provide strong evidence for the 1950 cohort, future research would benefit from linking draft lottery data with individual military service records and mortality information, where feasible.

\clearpage

\bibliographystyle{aea}
\bibliography{literature}

\clearpage

\appendix

\begin{landscape}
\section{Appendix} \label{Appendix}

\begin{table}[htbp]
\caption{Effect of draft-eligibility on long-term mortality, cohorts 1944–1952}
\label{reg_results_multi}
\centering
\renewcommand{\arraystretch}{1.15}
\begin{tabular}{lccccccccc}

\toprule
 & 1944 & 1945 & 1946 & 1947 & 1948 & 1949 & 1950 & 1951 & 1952 \\
\midrule
\multicolumn{10}{l}{\textbf{Panel A: Men only}} \\
Black              & -0.0037  & -0.0008  & 0.0036   & 0.0015   & 0.0007   & -0.0002   & 0.0051   & 0.0007 & -0.0014 \\
                   & (0.0029) & (0.0030) & (0.0030) & (0.0029) & (0.0029) & (0.0029)  & (0.0029) & (0.0029) & (0.0027) \\[0.15cm]
\midrule
Sample size     & 188{,}205 & 170{,}471 & 185{,}124 & 192{,}153 & 175{,}168 & 168{,}541 & 162{,}974 & 162{,}567 & 159{,}303 \\
\midrule

\multicolumn{10}{l}{\textbf{Panel B: Men \& women}} \\
Black              & 0.0031   &  0.0027  & -0.0034  & 0.0088   & 0.0002    & 0.0037    & -0.0058  & -0.0054 & -0.0030 \\
                   & (0.0038) & (0.0040) & (0.0039) & (0.0038) & (0.0039)  & (0.0039)  & (0.0040) & (0.0040) & (0.0037) \\
Men                & 0.0003   & 0.0008   & -0.0012  & 0.0013   & -0.0001   & -0.0015   & -0.0027  & -0.0026 & 0.0026 \\
                   & (0.0020) & (0.0021) & (0.0020) & (0.0020) & (0.0022)  & (0.0022)  & (0.0023) & (0.0023) & (0.0021) \\
Black × Men        & -0.0068  & -0.0035  & 0.0071   & -0.0073  & 0.0006    & -0.0038   & 0.0109   & 0.0061 & 0.0016 \\
                   & (0.0048) & (0.0050) & (0.0049) & (0.0048) & (0.0048)  & (0.0049)  & (0.0049) & (0.0049) & (0.0045) \\

\midrule
Sample size & 297{,}822 & 268{,}230 & 289{,}512 & 299{,}377 & 270{,}326 & 256{,}722 & 245{,}151 & 241{,}353 & 235{,}473 \\
\bottomrule
\end{tabular}

\begin{tablenotes}
Estimates from OLS regression with controls for month of birth fixed effects. Robust standard errors are reported in parentheses.
\end{tablenotes}

\end{table}
\end{landscape}

\end{document}